\documentclass[usenatbib]{mnras}

\usepackage{graphicx}	
\usepackage{amsmath}	
\usepackage{amssymb}	

\title[Young stars raining: PW~1]{Young stars raining through the Galactic Halo: the nature and orbit of 
Price-Whelan~1}

\author[M. Bellazzini et al.]{
Michele Bellazzini$^{1}$\thanks{E-mail: michele.bellazzini@inaf.it},
Rodrigo A. Ibata$^{2}$,
Nicolas Martin$^{2,3}$,
Khyati Malhan$^{4}$,\newauthor
Antonino Marasco$^{5,6}$
and Benoit Famaey$^{2}$,
\\
$^{1}$INAF - Osservatorio di Astrofisica e Scienza dello Spazio, via Gobetti 93/3, 40129 Bologna, Italy\\
$^{2}$Observatoire Astronomique de Strasbourg, Universit\'e de Strasbourg, CNRS UMR7550, Strasbourg, France\\
$^{3}$Max-Planck-Institut fur Astronomie, Heidelberg, Germany\\
$^{4}$The Oskar Klein Centre for Cosmoparticle Physics, Department of Physics, Stockholm University, AlbaNova, Stockholm, Sweden\\
$^{5}$Kapteyn Astronomical Institute, University of Groningen, PO Box 800, 9700 AV Groningen, The Netherlands\\
$^{6}$INAF - Osservatorio Astrofisco di Arcetri, largo E. Fermi 5, 50127 Firenze, Italy
}

\date{Accepted 2019 October 1. Received 2019 October 1; in original form 2019 July 30}

\pubyear{2019}

\begin{document}
\label{firstpage}
\pagerange{\pageref{firstpage}--\pageref{lastpage}}
\maketitle

\begin{abstract}
We present radial velocities for five member stars of the recently discovered young (age$\simeq 100-150$~Myr) stellar system Price-Whelan~1 (PW~1), that is located far away in the Galactic Halo (D$\simeq 29$~kpc, Z$\simeq 15$~kpc), and that is { probably} associated to the Leading Arm (LA) of the Magellanic Stream. We { measure} the systemic radial velocity of PW~1, $V_r=275 \pm 10$~km/s, significantly larger than the velocity of the LA gas in the same direction. We re-discuss the main properties and the origin of this system in the light of these new observations, computing the orbit of the system and comparing its velocity with that of the \ion{H}{i} in its surroundings. We show that the bulk of the gas at the velocity of the stars is more than $10\degr$ (5~kpc) away from PW~1 and the velocity difference between the gas and the stars become larger as gas closer to the stars is considered. { We discuss the possibilities that (a) the parent gas cloud was dissolved by the interaction with the Galactic gas, and (b) that the parent cloud is the high velocity cloud HVC~287.5+22.5+240, lagging behind 
the stellar system by $\simeq 25$~km/s and $\simeq 10\degr\simeq 5$~kpc. This HVC, that is part of the LA, has metallicity similar to PW~1, displays a strong magnetic field that should help to stabilise the cloud agains ram pressure, and shows traces of molecular hydrogen.} We also show that the system is constituted of three distinct pieces that do not differ only by position in the sky but also by stellar content.  
\end{abstract}




\section{Introduction}
\label{intro}

The second data release (DR2) of the Gaia space mission \citep[see, e.g.,][and references therein]{gaia_bro,gaia_lin} has provided an unprecedented view of our own Galaxy, leading to the discovery of
several new substructures and stellar systems of different provenance and nature 
\citep[][among others]{gaia-ence,ghostly,gap_abyss,CG18,CG19,antlia2}. While those studies used a variety of search techniques, an obvious and simple way to look for new stellar systems is to inspect density maps of blue stars. For instance, selecting stars with $(BP-RP)_0<0.0$ \citep[see][for a description of the Gaia DR2 photometric system]{gaia_eva} implies picking out mainly populations younger than $\simeq 500$~Myr, tracing relatively recent star formation, and blue horizontal branch (BHB) stars, tracing old and metal-poor stars, while removing from the sample the vast majority of Galactic stars \citep{gaia_babu}.

\begin{figure*}
	\includegraphics[width=\textwidth]{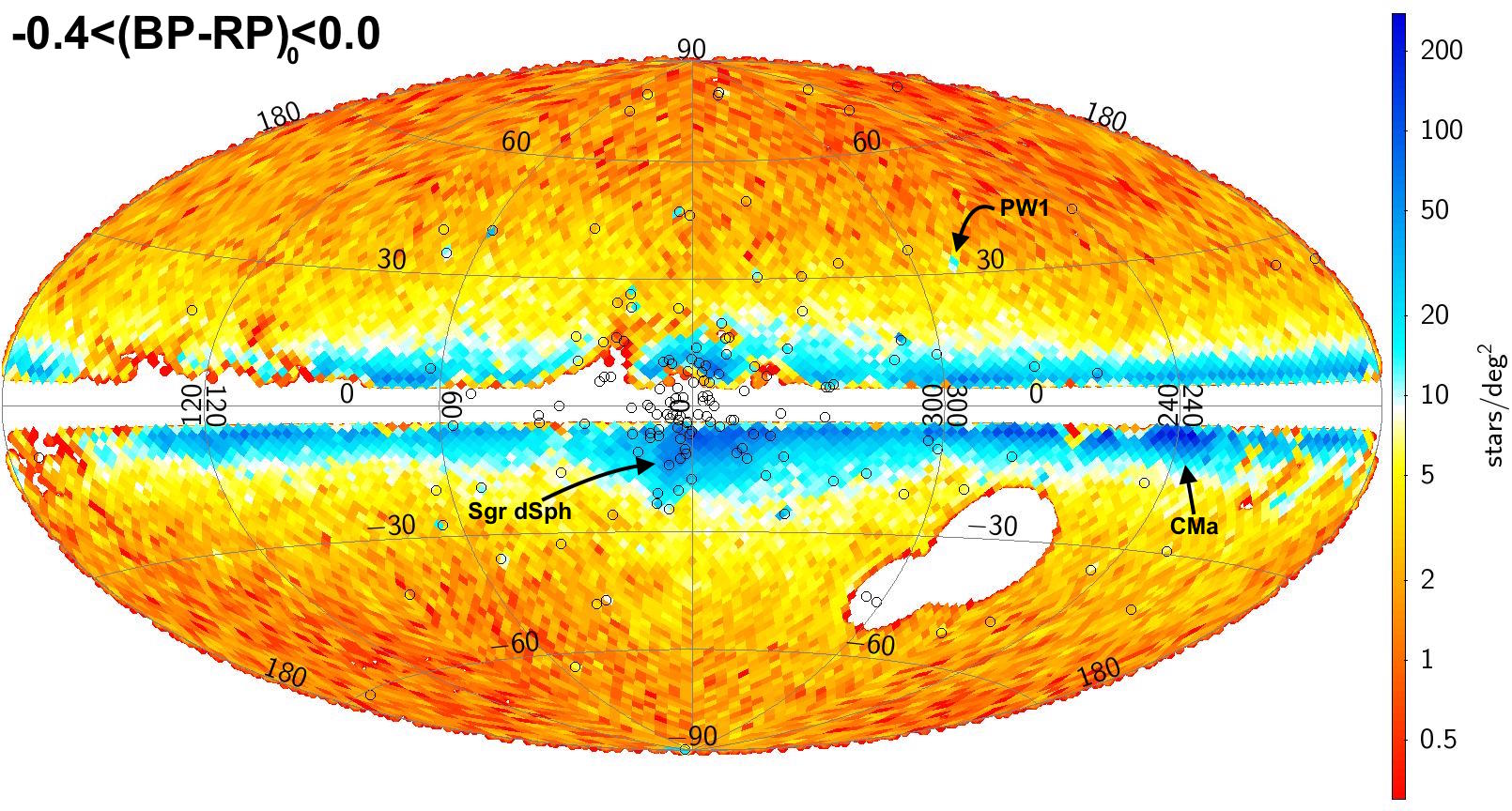}
    \caption{ HEALPix density map (in logarithmic scale) of blue stars with $\lvert b\rvert>5.0\degr$ from Gaia DR2 in Aitoff projection (Galactic coordinates).  Two circular regions including the Magellanic Clouds have been excised. Open circles are the Galactic globular clusters from the 2010 version of the catalog by \citet{h96}.  The position of Sgr~dSph and the Canis Major over-density are indicated and labelled. }
    \label{largemap}
\end{figure*}

A good example of the potential of this approach is provided by the map shown in Fig.~\ref{largemap}. The map has been obtained, from Gaia DR2 data, from a first selection in observed color\footnote{Not corrected for interstellar extinction.} $(BP-RP)<0.5$, in absolute Galactic latitude $\lvert b\rvert>5.0\degr$, in {\tt phot\_bp\_rp\_excess\_factor}, according to Eq.~C.2 by \citet{gaia_lin}, and keeping only sources with uncertainties in proper motions $<1.0$~mas/yr in both directions\footnote{These selections on the {\tt phot\_bp\_rp\_excess\_factor} and on the uncertainties in the proper motions are applied to all the samples extracted from Gaia DR2 that are used in this paper.}.
These selections lead to a sample of 2175166 stars for which we obtain an estimate of the interstellar reddening E(B-V) from the \citet{sfd} maps, corrected with the calibration by \citet{sf11}. Then we excised from the sample two wide circles around the position of the Large Magellanic Cloud (LMC) and Small Magellanic Cloud (SMC).
The resulting map in Fig.~\ref{largemap} displays the 233698 stars, from the sample described above, satisfying the condition $-0.4<(BP-RP)_0<0.0$. 

The most prominent structure in the map is the portion of the Galactic disc emerging at all longitudes just above our latitude limit $\lvert b\rvert=5.0\degr$, where stars younger than $\simeq 500$~Myr are very abundant. Within the disc the most remarkable substructure is the Canis Major over-density \citep[CMa,][]{mart04,bellaz06,moma}, traced by its young main sequence stars \citep[Blue Plume, see][]{butler}. On the other hand the galactic Bulge is traced by its conspicuous population of BHB stars \citep[see, e.g.,][and references therein]{bulgehb}. The same happens for the Sagittarius dwarf spheroidal \citep[Sgr dSph,][]{sgr0,m03,sgrbhb}, whose elongated shape is seen to plunge almost perpendicularly into the Galactic disc from the southern half of the Galaxy. Above and below the plane several small-scale over-densities are clearly visible as cyan-blue dots. Virtually all of them, { especially at high Galactic latitudes}, are Galactic globular clusters (GGCs) that possess a significant BHB population, as demonstrated by their match with the positions of GGCs from the \citet{h96} catalog (black open circles). { The only remarkable exception 
at $\lvert b\rvert\ge 15.0\degr$ is the isolated cyan dot at (l,b)=($288.5\degr, 31.9\degr$).}

This small-scale over-density turned out to have no known counterpart in catalogues of star clusters and/or nearby dwarf galaxies, it is an obvious clump also in the distribution of proper motions, and its Color Magnitude Diagram (CMD) reveals the presence of a prominent main sequence typical of a young stellar cluster located at a large distance ($> 20$~kpc), far in the Galactic Halo. While trying to understand the nature of this strikingly unusual object we realised that it had already been found, essentially with the same technique, by 
\citet[][P19, hereafter]{pw1}, who reached conclusions very similar to those independently reached by us.
In the following, we adopt their nomenclature, if not otherwise stated, starting from the name of the system:
Price-Whelan~1, hereafter for brevity, PW~1.

According to the analysis by \citet{pw1}, PW~1 is made up of a group ($M_{*}\simeq 1200$~M$_{\sun}$) of mildly metal-poor ([Fe/H]$\simeq -1.1$), $130\pm 6$~Myr old stars, located at a distance of $D=28.9\pm 0.1$~kpc, with an
interstellar extinction of $A_V=0.33\pm0.02$.
These values come from a the statistical analysis of a deep CMD obtained from DECam images. 
The degeneracy between $A_V$, distance and age leaves room for uncertainties on these parameters that may be significantly larger than those reported by P19. For instance, their estimate of the interstellar extinction toward PW~1 is $\simeq 0.1$~mag larger than what is obtained from the \citet{sfd} maps, corrected following \citet{sf11}.
Still, the conclusion that PW~1 is the only stellar system younger than 1~Gyr known to inhabit the Galactic Halo, at a distance of about 30~kpc from us, is very robust (see below for further analysis and discussion). In the following we will adopt the distance and metallicity estimates from P19, keeping in mind that the constraints on metallicity coming from the CMD are quite coarse. We also adopt the following reddening laws, provided by the on-line tool for the PARSEC stellar models \footnote{CMD 3.3 \tt http://stev.oapd.inaf.it/cgi-bin/cmd} \citep[see, e.g.][]{bressan,marigo}: $A_G=0.859E(B-V)$, $A_{BP}=1.068E(B-V)$, $A_{RP}=0.652E(B-V)$, for the Gaia DR2 photometry, and $A_g=1.165E(B-V)$, $A_i=0.676E(B-V)$, for PanSTARRS photometry \citep{pan}.
Throughout the paper we use only PARSEC stellar models \citep{bressan} produced with the same tool, adopting the default configuration for the initial mass function and the bolometric corrections.

The spatial distribution of the stars in PW~1 does not show the radial symmetry typical of star clusters and dwarf galaxies. The stars are located in at least two main pieces, according to the nomenclature by P19: "a", a sparse main body containing most of the system's stars, and "b", a minor but more compact component lying to the North of "a", separated by $\simeq 1.9\degr \simeq 1.0$~kpc. The internal motions are unresolved and P19 estimate a collective systemic motion in the plane of the sky of $\langle \mu_{\alpha}\rangle=-0.56\pm 0.04$~mas/yr\footnote{Where $\langle \mu_{\alpha}\rangle$ must be intended as $\langle {\rm \mu_{\alpha}cos({\delta})}\rangle$.}and $\langle {\rm \mu_{\delta}}\rangle=+0.47\pm 0.02$~mas/yr. According to P19, this newly-formed stellar system is likely slowly dissolving, as suggested by its sparse nature and odd morphology.

PW~1 is projected near the edge of one of the lanes (L~II) of the so-called Leading Arm \citep[LA,][]{mary}, the { arm} of the Magellanic Stream \citep[MS, see P19,][and references therein]{nide10,dong} that leads the Magellanic Clouds beyond the Galactic plane \citep[see also][]{nide08,nide10,venz}. { The very existence of a leading arm is interpreted as an evidence that Galactic tides had a major role in the origin of the MS \citep[see, e.g.,][for a thorough discussion of recent models for the formation of the MS]{dong}. However, other hypotheses are also being considered, as, for instance, that the LA is formed from ram-pressure stripping of a satellite of the Magellanic Clouds that is moving ahead of them \citep{hammer,wang,tepper}.}
 
Assuming values of the radial velocity matching those of the main features of the LA lying near PW~1, P19 show that the time since the system crossed the Galactic disc is similar to the age of its stars. Based on this finding, they suggest that the compression suffered by the gas in the MS while crossing the Galactic Disc lead to the small star formation episode that gave birth to PW~1 as a stellar system. Once born, the stars become kinematically decoupled from the gas (as they do not suffer from the drag exerted by Galactic gas), following a slightly different orbit. 
This hypothesis is very intriguing and, indeed, provides a reasonable explanation of the available observational data  {\citep[including the similarity in the mean metallicity between PW~1 and L~II, see, e.g., P19][]{lu,wakker,fox,richter}}. If confirmed, it would have implications for, e.g., providing the means for a direct estimate of the distance to the Magellanic Stream and to probe the gas density in the hot circum-Galactic corona (see P19 for a detailed discussion of the relationship of PW~1 with the MS and with the Magellanic system as a whole).

In this contribution we present the first { measurement} of the heliocentric line-of-sight velocity (hereafter radial velocity, $V_r$) of PW~1, the only missing piece of the 3-D motion of this object, allowing us to compute an orbit fully based on observational data. Moreover, we reconsider the main properties of the system, showing that it is likely made of three, partially independent pieces that differ not only in position but also in the mass range of their member stars. 

\section{The structure of PW~1}
\label{structure}

\begin{figure}
	\includegraphics[width=\columnwidth]{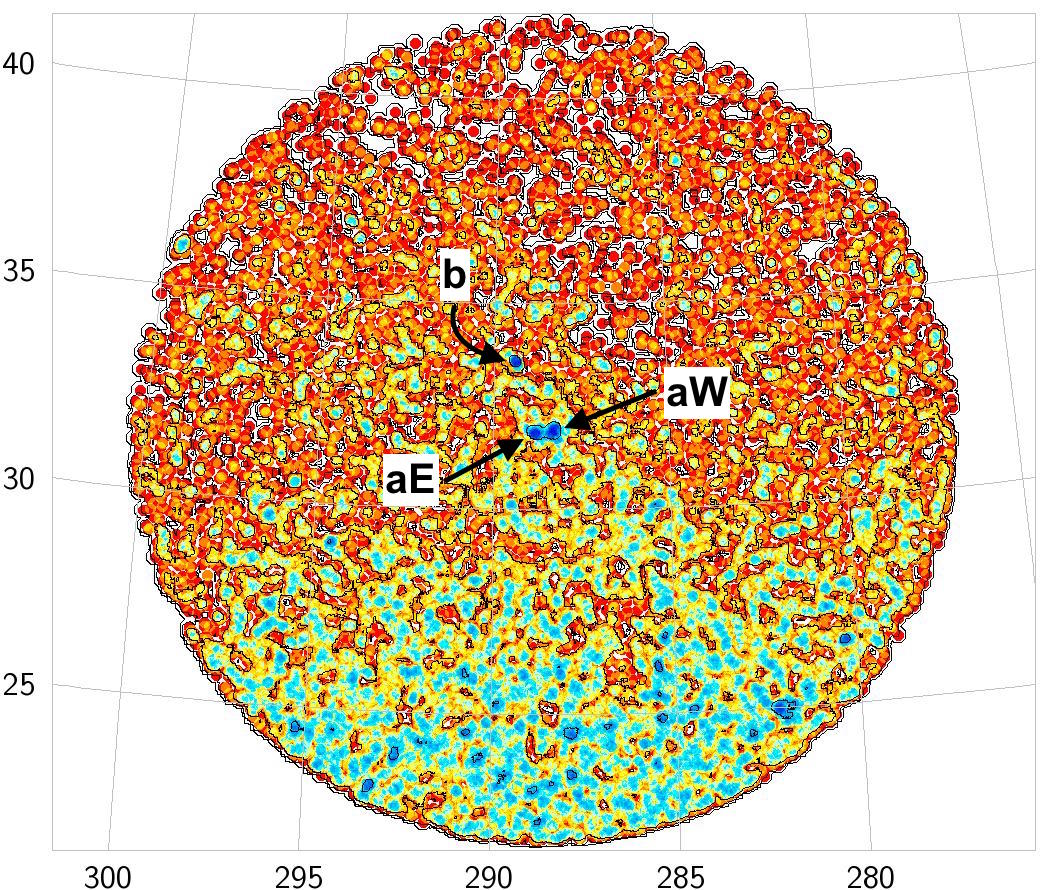}
    \caption{Density map (Galactic coordinates, logarithmic scale) of stars from Gaia DR2 in the region surrounding PW~1.
    Stars in this map are selected to have proper motion within 1.0~mas/yr from the mean motion of PW~1 and in parallax, to reject obvious nearby stars. The three components of the stellar system are indicated by arrows and labelled.  }
    \label{smallmap}
\end{figure}
\begin{figure}
	\includegraphics[width=\columnwidth]{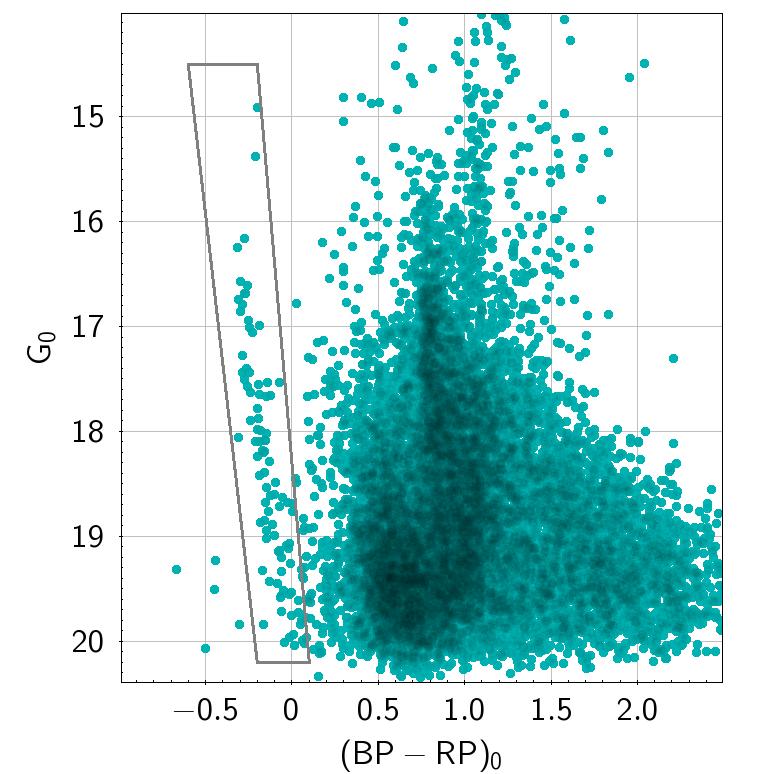}
    \caption{Color magnitude diagram of the same parallax and proper motion selected stellar sample shown in Fig.~\ref{smallmap}. The polygon shows the box in which we select the blue MS stars that are the 
    characteristic population of PW~1.}
    \label{bluecmd}
\end{figure}

In Fig.~\ref{smallmap} we show a density map of stars from Gaia DR2 lying within $10\degr$ from the center of PW~1, with a magnitude-dependent selection in parallax aimed at rejecting stars that are clearly at a lower distance with respect to PW~1\footnote{In particular we retain in the sample only stars having $\lvert {\tt parallax}\rvert < 3.0(-16.4230 +3.3644*G_0 -0.2273G_0^2 +0.0051G_0^3)$. Note that proper motion and color selected PW~1 members form a wedge-shaped distribution around ${\tt parallax}=0.0$~mas in the parallax vs. $G_0$ plane. The wedge is narrow at bright magnitudes and widens toward the faint end, as the errors in parallax become larger with magnitude.}, and having proper motion within $1.0$~mas/yr from the mean motion of the system. Note that $1.0$~mas/yr corresponds to the typical uncertainty on proper motion for the faintest stars in our sample. Hence, while it may appear excessively relaxed, as it corresponds to $\simeq 137$~km/s at D=28.9~kpc, it is conservatively set to minimize the loss of actual members, especially at faint magnitudes. This is particularly convenient, in the present case, since the color of PW~1 stars within the reach of Gaia DR2 is so blue that, when proper color-cuts are adopted, the contamination from fore/background stars is negligible. In any case, all the results presented below are fully confirmed also when more stringent selection criteria are adopted (e.g., retaining only stars with proper motion within $0.5$~mas/yr from the mean motion of PW~1). 

A large scale density gradient is clearly visible in Fig.~\ref{smallmap}, as the Galactic disc, that dominates the surface density at 
$b\la 30\degr$, gently disappears going toward the Northern Galactic Pole. PW~1 appears as three distinct compact density clumps. The main body "a", at the center of the map, is split into an Eastern and a Western component, aE and aW, hereafter. There is a hint of a diagonal band, emanating from the edge of the disc at (l,b)$\simeq(286\degr,28\degr)$ and reaching (l,b)$\simeq(300\degr,34\degr)$, that encloses PW~1. However the stellar population and kinematics of the stars in this band are indistinguishable from those in the fore/background population surrounding PW~1, thus are likely not related to it. 

The CMD of all the stars displayed in this map is shown in Fig.~\ref{bluecmd}. The narrow blue main sequence
of PW~1 is clearly evident, emerging at $(BP-RP)_0\la 0.1$ and $G_0\la 20.0$ from the typical Disc and Halo fore/background population (see P19 for the comparison with the CMD of control fields). In this plot we also show the box by which we further select the most likely members of PW~1 in color and magnitude. 

\begin{figure}
	\includegraphics[width=\columnwidth]{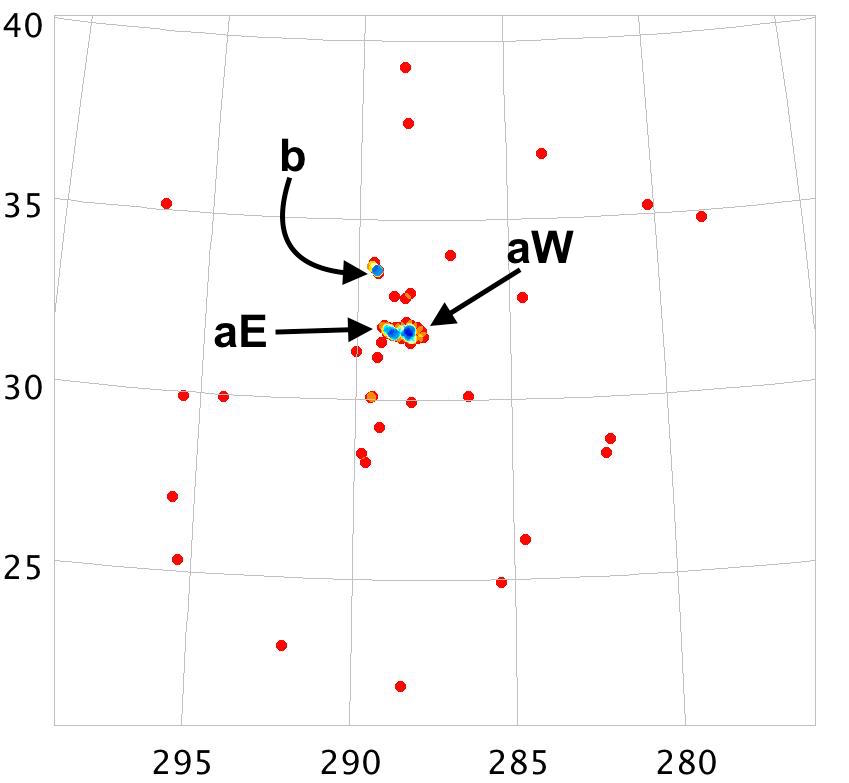}
    \caption{Density map (Galactic coordinates, logarithmic scale) of stars from the blue MS of PW~1, selected with the color-magnitude box shown in Fig.~\ref{bluecmd}. 
    }
    \label{bluemap}
\end{figure}

The stars selected with this box are displayed in Fig.~\ref{bluemap}. This further selection virtually removes everything not related to PW~1. It would be very interesting to follow up spectroscopically all these stars since, if they are confirmed as members, they may be the best witnesses of the likely on-going dissolution of the system. However, it would be surprising if the three stars lying in between aE+aW and b as well as those forming a kind of tail from the Southern edge of aE to $b\simeq -33\degr$, running nearly parallel to the $b=180\degr$ line were not (previous) members of the system. This map suggests that in the surrounding of PW~1
there are several blue MS stars likely related to it.

\begin{figure}
	\includegraphics[width=\columnwidth]{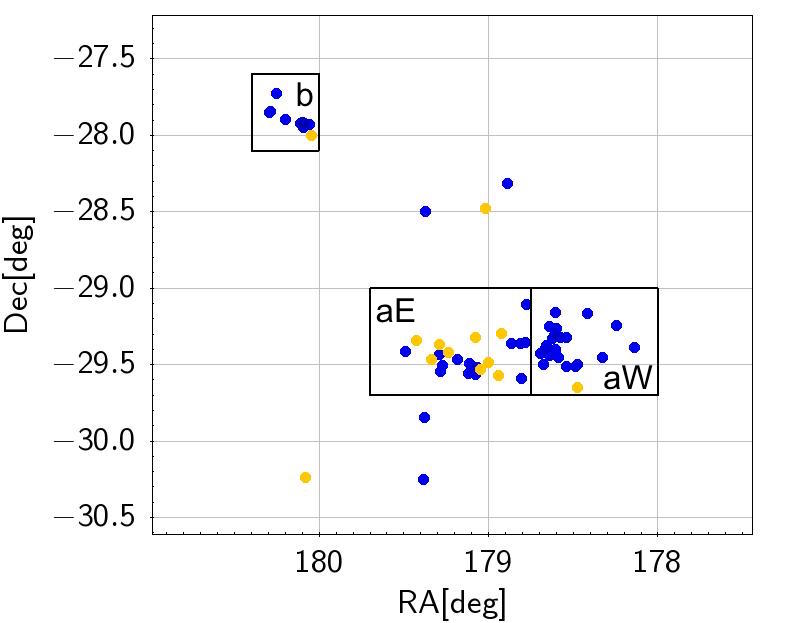}
    \caption{Zoomed version of Fig.~\ref{bluemap} with boxes selecting stars in the three different pieces of PW~1. Yellow points mark stars having $G_0<17.1$.
    }
    \label{trepezzi}
\end{figure}

Fig.~\ref{trepezzi} provides a zoomed view of the same map, with the boxes we will adopt in the following to attribute stars to the various pieces of PW~1 (aE, aW, b). { Triggered} by the gap in the blue MS appearing at $G_0= 17.1$ in the CMD of Fig.~\ref{bluecmd}, here we plot in yellow the stars brighter than this limit. It is interesting to note that only one of these fifteen bright stars is associated to aW, according to our selection, and this is the faintest one. aE and aW have the same total number of attributed members (25) but in aE nine of them are brighter than the apparent gap in the blue MS. One (of 25), as said is in aW and one (of 13) is in b. The remaining four bright stars are spread around the surrounding area. While its is hard to establish the statistical significance of this difference in the content of bright stars between aE and aW, Fig.~\ref{trepezzi} suggests that some degree of mass segregation is present within PW~1. We will discuss this feature in more detail  in Sec.~\ref{discu}, below.

\section{Spectroscopic observations}
\label {obs}

\begin{figure}
	\includegraphics[width=\columnwidth]{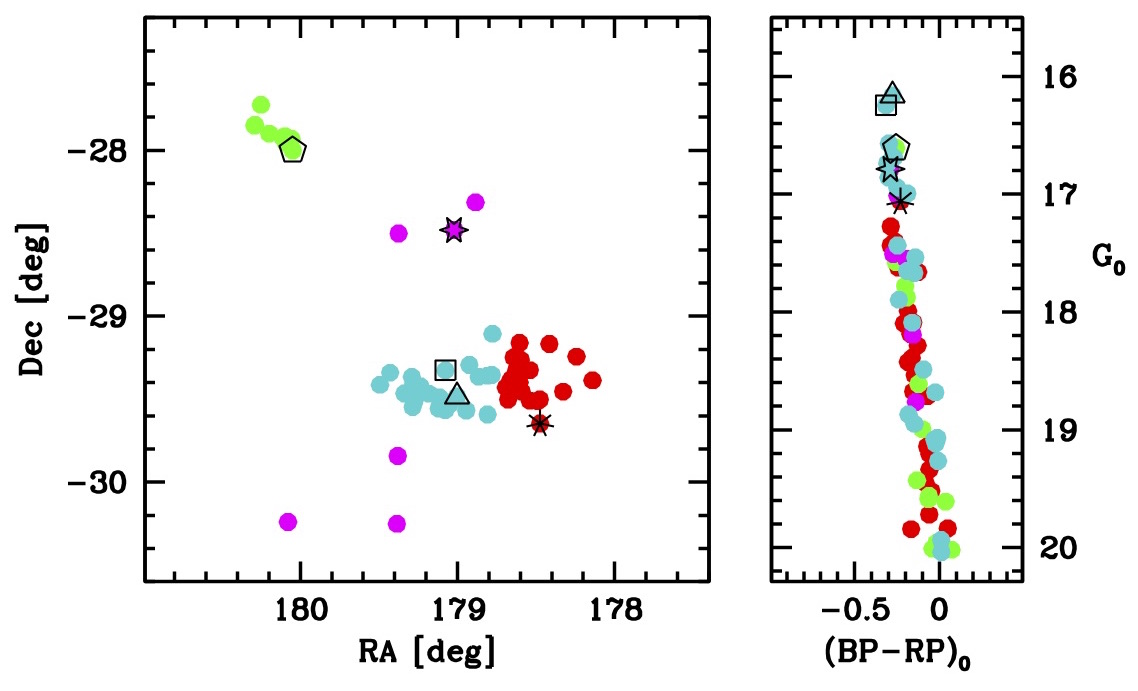}
    \caption{Positions of the spectroscopic targets in the sky map (left panel) and in the CMD (right panel) of PW~1. The targets are plotted with the same empty symbols in both panels. In { both panels} the filled circles representing the stars of PW~1 are coloured according to the sub-system they belong to: aE (cyan), aW (red), b (green). The stars that are likely members of the system but cannot be attributed to any of these pieces are plotted in magenta. 
    }
    \label{maptarg}
\end{figure}

\begin{table*}
	\centering
	\caption{Heliocentric radial velocities and other properties of the target stars}
	\label{tab1}
	\begin{tabular}{lcccccccc} 
		\hline
name & RA & Dec & G$_0$ & BP$_0$ & RP$_0$ & E(B-V) & V$_r$ & Gaia DR2 id \\
     &[deg]&[deg]&[mag]&[mag]&[mag]&[mag]&[km/s]& \\ 
		\hline
 aE1 & 179.00182010 & -29.48336292 & 16.161  & 16.015 &  16.290 & 0.054 & $273.8\pm 11.1$ &  3480054567924428032 \\
 aE2 & 179.07677448 & -29.32545291 & 16.245  & 16.076 &  16.390 & 0.056 & $271.3\pm 13.0$ &  3480064910205689088 \\
  b1 & 180.04853086 & -28.00034847 & 16.608  & 16.481 &  16.736 & 0.064 & $297.5\pm 19.7$ &  3486242378847130624 \\
  m1 & 179.02061045 & -28.47964538 & 16.787  & 16.634 &  16.923 & 0.062 & $253.1\pm 25.9$ &  3486167027940903296 \\
 aW1 & 178.47412626 & -29.64657881 & 17.060  & 16.929 &  17.160 & 0.043 & $280.4\pm 28.4$ &  3480036975738300800 \\
\hline
	\end{tabular}
\end{table*}

We obtained optical spectra of five blue main sequence stars from the central part of PW~1 selected as illustrated in Fig.~\ref{maptarg}. We choose two stars in aE (that we name aE1 and aE2, for brevity, hereafter; open triangle and square in Fig.~\ref{maptarg}), one from aW (aW1, asterisc), one from b (b1, open pentagon) and one from the group of three stars lying between aE+aW and b (m1, open star). The spectrum of a sixth star (belonging to aW) was also acquired but the signal-to-noise was too low to be useful, and we excluded it from further analysis. The main properties of these stars, including their Gaia DR2 identificative number, are reported in Tab.~\ref{tab1}.

Observations were obtained during the nights of June 6 and 7, 2019, using the spectrograph EFOSC2 mounted at the 3.5~m New Technology Telescope (NTT, ESO, La Silla). We took one $t_{exp}=900$~s spectrum per target using the Grism 
\#19 with the {\tt slit\#0.5\_red} slit, which is $0.5\arcsec$ but shifted with respect to the center of the EFOSC2 field so as to allow observation at redder wavelengths. This set-up covers the spectral range 4441\AA-5114\AA ~including the Mg`b' triplet, with a resolution
$\lambda/\Delta\lambda\sim 6500$ at the wavelength of the H$_{\beta}$ line. Observations of an Argon-Helium arc-lamp were taken immediately after every target star before moving the telescope to minimize uncertainties from instrument flexure.
All the spectra were reduced using standard IRAF procedures, including bias subtraction, flat-fielding, extraction, and wavelength calibration.

\begin{figure}
	\includegraphics[width=\columnwidth]{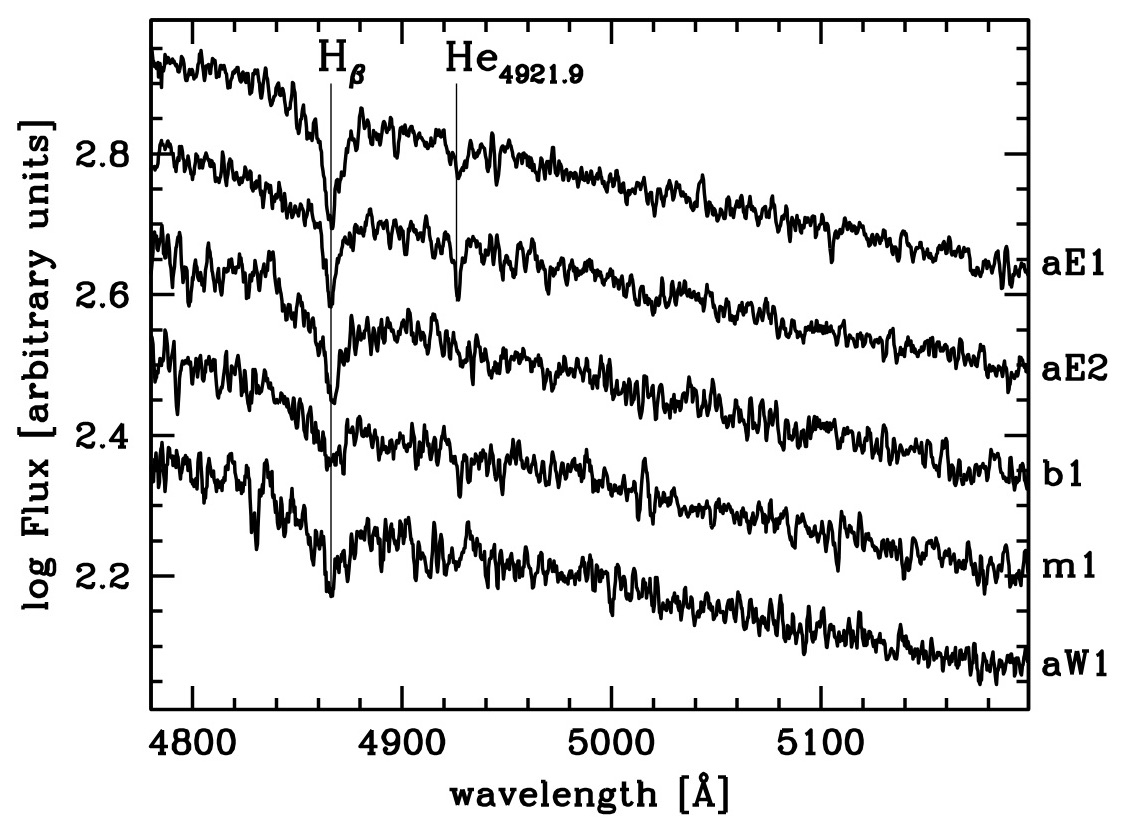}
    \caption{Spectra of the observed targets. Arbitrary shifts in log flux have been added to avoid overlapping the spectra within the panel. The only identifiable lines are marked and labelled.
    }
    \label{spectra}
\end{figure}

The final spectra are shown in Fig.~\ref{spectra}.
They all have similar signal to noise ratios (S/N$\sim 6-7$) but they differ in the strength of the only strong line that can be identified in all of them, that is $H_{\beta}$. Those having the strongest $H_{\beta}$, the two brightest and bluest stars aE1 and aE2, show also a hint of the { neutral} Helium line at 4921.9\AA. In the following analysis we will use only $H_{\beta}$ to estimate the radial velocity but we verified that the results obtained with the \ion{He}{i}4921.9 line for these two stars are fully consistent with those obtained with $H_{\beta}$, albeit with a higher uncertainty.

We used IRAF/splot to simultaneously fit the $H_{\beta}$ absorption line with a Lorentzian profile, and the adjacent continuum with a straight line. The line centroid positions and the associated uncertainties were propagated into observed radial velocities and, finally, into heliocentric radial velocities by applying the heliocentric corrections computed with IRAF/rvcorr. The final uncertainties range from $\sim 10$~km/s to $\sim 30$~km/s, depending on the strength of $H_{\beta}$ in the considered spectrum. This precision is clearly not sufficient to resolve the internal motions of PW~1 but is more than sufficient to obtain the first { measurement} of its systemic radial velocity, a crucial piece of information that was still missing (P19).

The observed velocities ranges from $V_r=253.1\pm 25.9$~km/s to $V_r=297.5\pm 19.7$~km/s. 
The comparison with the predictions of the Besancon Galaxy Model 
\citep{besa03}\footnote{\tt https://model.obs-besancon.fr} toward this line of sight strongly suggest that none of the observed stars can be ascribed to the Milky Way galaxy. In particular, in a $1.0$~deg$^2$ sample extracted from the model: (a) in the color range of our target stars no model source has an heliocentric radial velocity larger than $V_r=74.6$~km/s, and, (b) even considering stars of any color, only 9 over 4770 have $G<20.5$, $D>20$~kpc, and $V_r>+250$~km/s, and all of them have negative values of $\mu_{\delta}$. We conclude that all our target stars are members of PW~1. Therefore, the star m1, though it do not resides in any of the three main pieces of PW~1, is physically associated to the system.

\begin{figure}
	\includegraphics[width=\columnwidth]{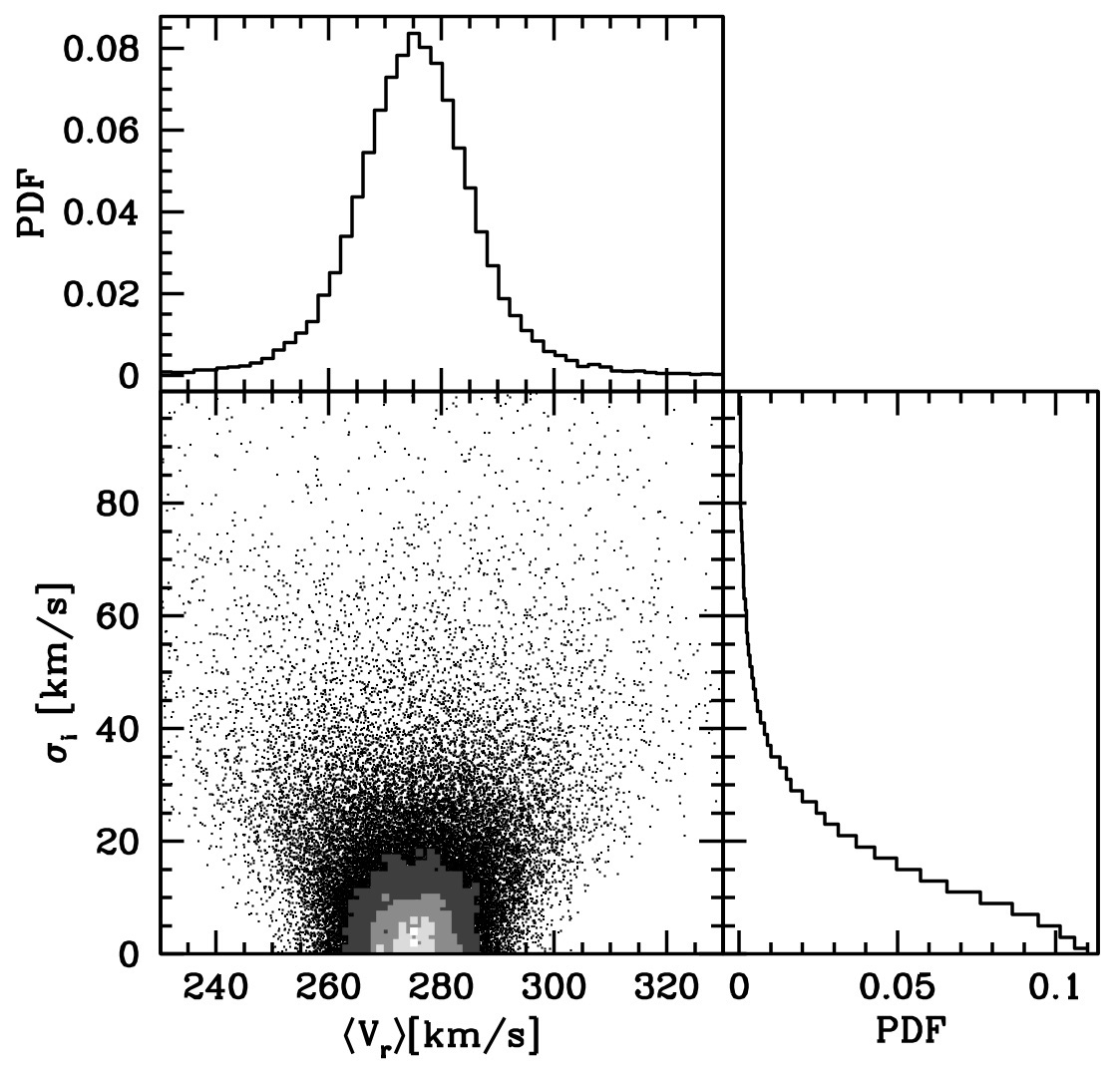}
    \caption{Joint PDF of the two-parameter Gaussian model (bottom left) and the marginalised PDF for the mean velocity (top) and the intrinsic velocity dispersion (right).}
    \label{chain}
\end{figure}

It may appear somehow intriguing that the two stars belonging to aE have very similar velocities ($273.8\pm 11.1$~km/s and $271.3\pm 13.0$~km/s), and different from the stars from other sub-systems of PW~1, but none of the observed differences is even marginally significant.
For this reason, we consider the five stars as tracers of the same velocity distribution, and we consequently estimated the mean velocity and intrinsic velocity dispersion, following \citet{mart18}. 
In practice, we infer the mean velocity $\langle V_r\rangle$ and the intrinsic velocity dispersion $\sigma_i$ by assuming a Gaussian distribution of the 5 velocity data points and uniform priors and by using the code 
JAGS\footnote{\tt http://mcmc-jags.sourceforge.net} within the R environment\footnote{\tt https://www.r-project.org}. The posterior PDF is shown in Fig.~\ref{chain}. The PDF of $\langle V_r\rangle$ is well behaved and Gaussian-like, hence, as our fiducial systemic velocity, we adopt the median of the PDF, with $1\sigma$ uncertainty given by the semi-difference between the 84th and 16th percentiles: $\langle V_r\rangle =275 \pm 10$~km/s (with $257\le \langle V_r\rangle\le 294$~km/s at the 90\% confidence level). This corresponds to 
$\langle V_{LSR}\rangle =272 \pm 10$~km/s in the Local Standard of Rest, and $V_{GSR}=84 \pm 10$~km/s in the Galactic standard of rest, adopting the solar motion by \citet{schon} and the circular speed of the Sun from \citet{McM}. In a frame aligned with the Galactocentric cylindrical coordinates, with $V_R$ pointing outwards from the Galactic center, $V_\phi$ in the direction of rotation and $V_z$ pointing to the North Galactic pole, the  3D velocity vector is $(V_R, V_\phi, V_z)\simeq (-16, 8, 188)$~km/s.
As expected, the intrinsic velocity dispersion is not resolved by our data. The mode of the PDF is $\sigma_i=0.0$~km/s, with $\sigma_i\le 32$~km/s at the 90\% confidence level.

It is very interesting to note that the derived radial velocity of PW~1 does not match the velocity of the nearby { clouds} of the LA. P19 identify three sub-structures as the possible gas counterparts of PW~1, at velocity $V_{LSR}\simeq$ 60, 110, and 230 km/s, all significantly lower than our estimate $\langle V_{LSR}\rangle =272 \pm 10$~km/s. 

\section{Discussion}
\label{discu}

\begin{figure}
	\includegraphics[width=\columnwidth]{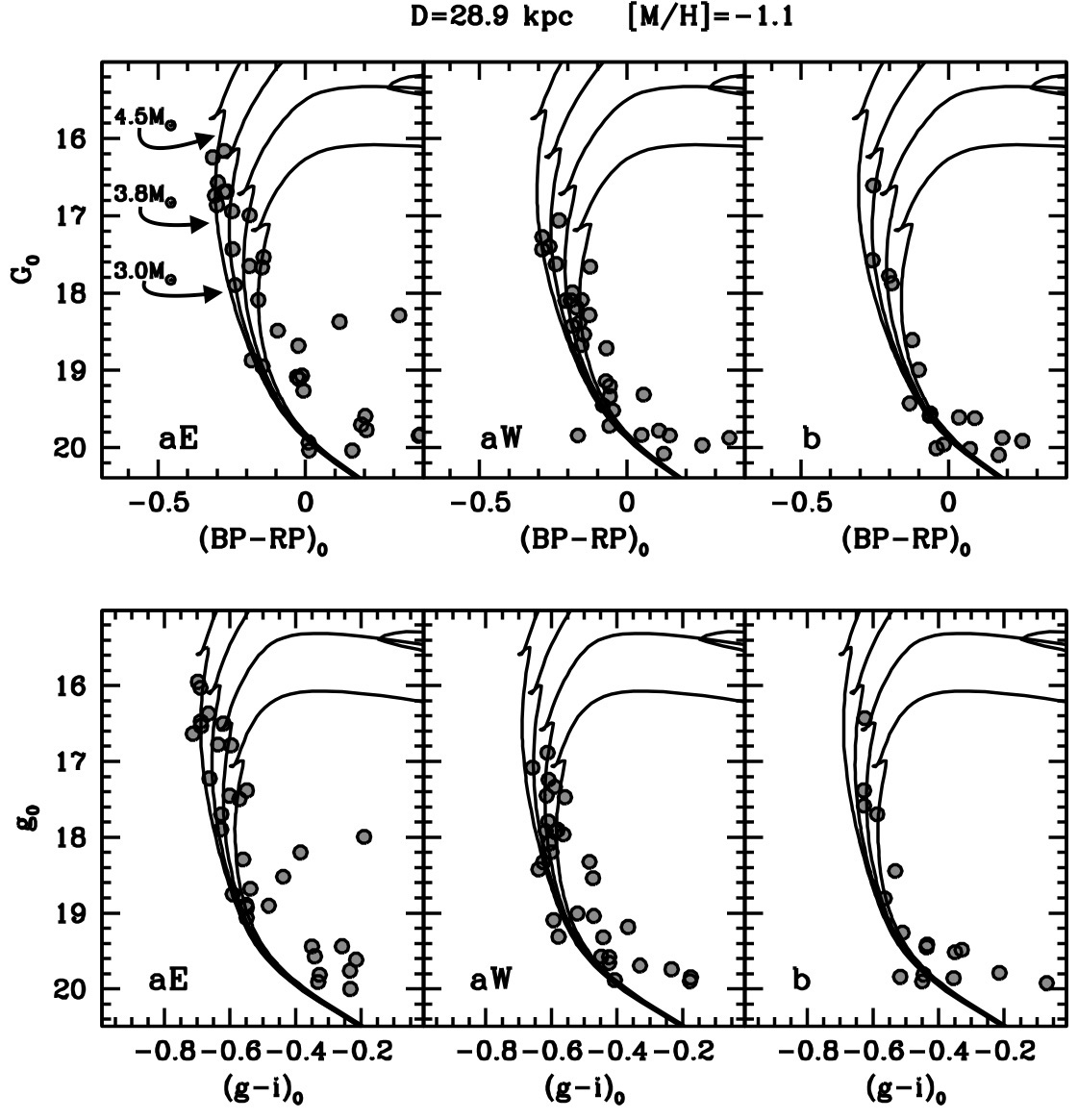}
    \caption{CMDs of the aE, aW and b subsystems (left, central, and right panels, respectively) from Gaia DR2 (upper row of panels) and PanSTARRS1 (lower row of panels) photometry. On each CMD we over-plotted four isochrones at metallicity [M/H]=-1.1, shifted to the distance of PW~1, with age$\simeq$ 100, 160, 250, and 400 Myr, { from left to right}, from the PARSEC set. In the upper left corner we indicated the initial stellar masses corresponding to $G_0=$16.0, 17.1, and 18.0 on the age=100~Myr isochrone.
    }
    \label{cmd}
\end{figure}

In Fig.~\ref{cmd} we compare the CMD of the three main pieces of PW~1 with a grid of four PARSEC isochrones, shifted to D=28.9~kpc, with [M/H]=-1.1 and age$\simeq$ 100, 160, 250, and 400 Myr.
The comparison is performed using both Gaia DR2 and Pan-STARRS1 magnitudes, as a consistency check. Only color-selected likely members are considered here. In the small portion of the CMD populated by PW~1 stars in Fig.~\ref{cmd}, the isochrones are nearly vertical, thus providing poor constraints on the distance. This is the main reason why we rely on the distance (and metallicity) estimate of P19, obtained from a deeper and more complete CMD, reaching $G_0\simeq g_0\ga 19.5$, where isochrones bend to the red, becoming more sensitive to this fundamental parameter.

The observed CMD is consistent with an age between 100 and 150~Myr, in agreement with the conclusions by P19. In this age regime and with such a sparse sample, the color of the brightest stars is the most sensitive and reliable age indicator. In this respect, Fig.~\ref{cmd} suggests that with these assumptions on distance and metallicity, an age younger than 100~Myr is unlikely.
It is also apparent from Fig.~\ref{cmd} that the CMDs of the three pieces are not identical. The most obvious difference is that virtually all the stars brighter than $G_0=17.1$ belong to aE, as anticipated in Sect.~\ref{structure}. If the distance, age and metallicity assumptions we made are correct, and we are dealing with genuine single stars, virtually all the stars with M$>3.8$~M$_{\sun}$, up to M$\simeq4.5$~M$_{\sun}$, are segregated into aE. This may be merely due to small number statistics, but also to slight differences in age, binary content and/or distance between 
the various pieces. For example, this correlation between position and stellar content may hint at a slightly different epoch of formation in different fragments of the parent gas cloud of PW~1. In any case it provides further support, in addition to the density field, to the idea that PW~1 is constituted by three distinct pieces.   

A rough estimate of the stellar mass of the system can be obtained by fitting a theoretical model to the observed  Luminosity Function (LF), thus allowing a sound extrapolation to the stars that went undetected because they are fainter than the limiting magnitude of our sample. To do this, we adopt a PARSEC model with age=100~Myr and [M/H]=-1.1, shifted to D=28.9~kpc, as above. To compare the observed and the model LFs in a fully homogeneous way, including the same binning, the theoretical LF was derived from a synthetic population of 12000~M$_{\sun}$ with the adopted age and metallicity, obtained with CMD~3.3.

\begin{figure}
	\includegraphics[width=\columnwidth]{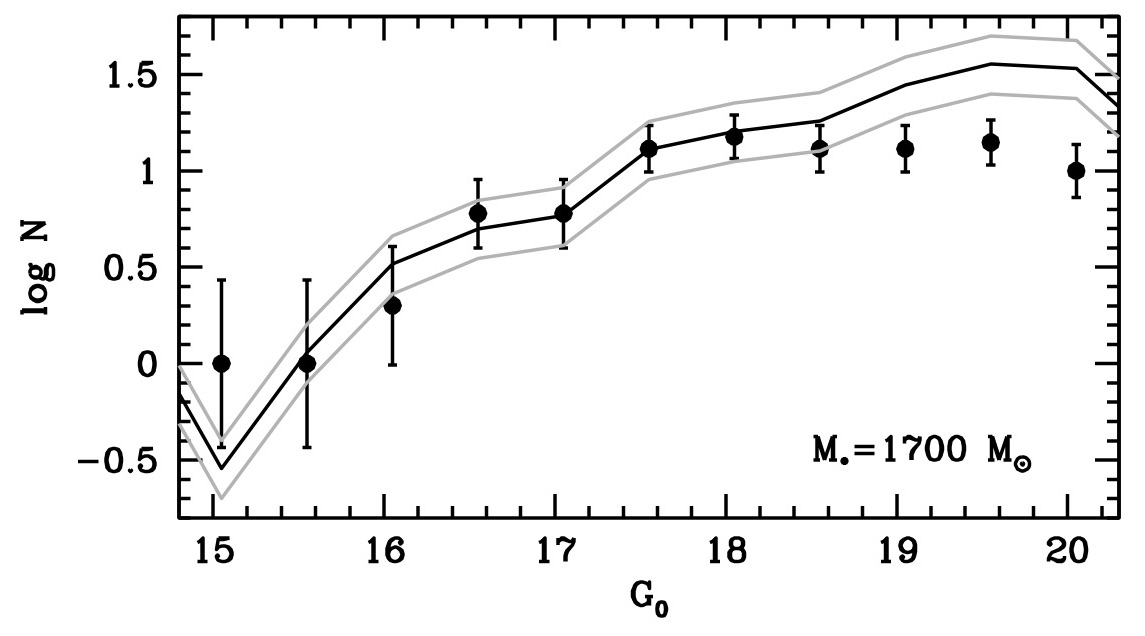}
    \caption{Luminosity Function (LF) of all the PM-selected, parallax-selected and color-selected possible members of PW~1 shown in Fig.~\ref{bluemap}. The black-line is the theoretical LF from an age=100~Myr, [M/H]=-1.1 PARSEC model that best-fits the observations in the range $G_0<18.5$, corresponding to M$_{*}\simeq 1700$~M$_{\sun}$. The grey curves are the theoretical LF adopting different normalisation factors, corresponding to M$_{*}=2000$~M$_{\sun}$ (upper curve) and to M$_{*}=1000$~M$_{\sun}$ (lower curve)
    }
    \label{lf}
\end{figure}

To obtain the best-fit normalisation between the observed and theoretical LFs, that corresponds to a total stellar mass, we minimize $\chi^2$ in the range $G_0<18.5$, as at fainter magnitudes the effect of incompleteness is apparent. In Fig.~\ref{lf} we show the best-fit to the LF of all the MS stars selected as possible current or former members of PW~1 (shown in Fig.~\ref{bluemap}), leading to M$_{*,tot}\simeq 1700$~M$_{\sun}$. We repeated the exercise for all the various pieces of PW~1 and we report the exact values of the best-fit normalizations in Tab.~\ref{tab2}, to be taken, as said above, with due caution, given all the systematic uncertainties that can affect the comparison. Reassuringly, for aE+aW we obtain M$_{*}\simeq 1100$~M$_{\sun}$, in good agreement with the results of P19 for the same part of the system (M$_{*}= 1200$~M$_{\sun}$). From the same model used to fit the LF we can obtain the mass-to-light ratio in the Johnson-Cousin V band ($M/L_V=0.13$), and, consequently, we convert the mass estimates into estimates of $M_V$, also reported in Tab.~\ref{tab2}.

Following \citet{CG18}, as a size indicator we computed r$_{50}$ (in arcmin, R$_{50}$ in parsec), that is the radius enclosing half of the members of each sub-system, adopting the memberships illustrated in Fig.~\ref{trepezzi}. This choice has two desirable features: first, r$_{50}$ is the best proxy for the half light radius that we can obtain from our data, second it allows a direct and fully homogeneous comparison with the 1229 Galactic open clusters for which \citet{CG18} provide an estimate of the same parameter. 
Adopting their modal distance ($D_{mod}$), the open clusters of \citet{CG18} have 
0.2$< R_{50} <$15.2~pc, with a median of 2.75~pc and 90\% of the sample having R$_{50}\le$ 5.2~pc. It is very interesting to note that the only piece of PW~1 that has a size comparable with that of star clusters is b (R$_{50}=26$~pc). On the other hand, aE, aW and the two taken together have R$_{50}\sim 100-150$~pc, in the range spanned by dwarf galaxies of similar luminosity \citep[see, e.g.,][]{mc12,antlia2}. As already noted by P19, PW~1 has the stellar mass and stellar population of an open cluster but the size of a dwarf galaxy, suggesting that we are seeing the system while it is dissolving into the halo, being not bound by self-gravity. It is important to stress that these estimates of r$_{50}$ were only intended to provide an idea of the typical size of the fragments. Given their irregular morphology and the lack of obvious centres of symmetry, a characteristic radius is clearly not fully adequate to describe the structure of PW~1 pieces. In particular aE is nearly filamentary, with a shell morphology that is observed also in other places where star formation is occurring, triggered by interactions \citep[e.g.,][]{david}.

If PW~1 is indeed the product of a small recent episode of star formation that occurred within a cold gas cloud, associated or not associated with the MS (see below) and likely triggered by interaction with Galactic gas, { an interesing} similarity, albeit at a smaller scale, can be noted with SECCO~1 \citep[see][and references therein]{alone}. This is a low mass ($M_{*}=10^5$~M$_{\sun}$) star-forming system (with no detection of stars older than $\sim 50$~Myr) located in the outskirts of the Virgo cluster of galaxies, that is the possible prototype of a new class of stellar systems that are born in isolated \ion{H}{i} clouds kept together by the external pressure of the surrounding hot gas \citep{sand17,alone}.  SECCO~1 is fragmented in two pieces, each one with size of a few hundreds of pc, and will probably fragment into smaller pieces in the near future. While the largest sub-system of SECCO~1 (main body, MB) lies within an \ion{H}{i} cloud of $\sim 10^7$~M$_{\sun}$, the smallest fragment (secondary body, SB) has no cold gas detected in coincidence with the stellar body, but it is likely associated with a $M\simeq 10^6$~M$_{\sun}$ cloudlet having the same velocity but off-set by $\sim 2.5$~kpc, in projection, with respect to the stars. We will see below that this is reminiscent of the configuration of PW~1 with respect to a nearby isolated small gas cloud.

\begin{table}
	\centering
	\caption{Structural parameters of PW~1 and its substructures.}
	\label{tab2}
	\begin{tabular}{lcccccc} 
		\hline
name & RA$_0^a$ & Dec$_0$ & M$_{*}^b$ & $M_V^c$ & r$_{50}^d$ & R$_{50}^e$  \\
     &[deg]&[deg]&[M$_{\sun}$]&[mag]&[arcmin]&[pc] \\ 
		\hline
PW~1 (all)&      &       & $\le 1716$ & $\ge-5.5$ &      &     \\
aE+aW  & 178.8& -29.4 & 1056 & -4.9 & 17.9 & 150 \\
aE     & 179.1& -29.5 &  684 & -4.5 & 14.1 & 118 \\
aW     & 178.6& -29.3 &  516 & -4.2 & 12.9 & 108 \\
b      & 180.1& -27.9 &  192 & -3.1 &  3.1 &  26 \\
\hline
\multicolumn{7}{l}{$^a$ RA$_0$, Dec$_0$ are the coordinates of the density peak of the subsystem,} \\
\multicolumn{7}{l}{except for aE+aW, for which we adopted the position by P19.}\\
\multicolumn{7}{l}{$^b$ Best fit values from fitting the LF as in Fig.~\ref{lf}}. \\
\multicolumn{7}{l}{$^c$ Adopting M/L$_V$=0.13, from the age=100~Myr and [M/H]=-1.1 model.} \\
\multicolumn{7}{l}{$^d$ Angular radius enclosing half of the members.} \\
\multicolumn{7}{l}{$^e$ Physical radius enclosing half of the members, adopting D=28.9~kpc.} \\
	\end{tabular}
\end{table}

\subsection{The orbit of PW~1}

\begin{figure}
	\includegraphics[width=\columnwidth]{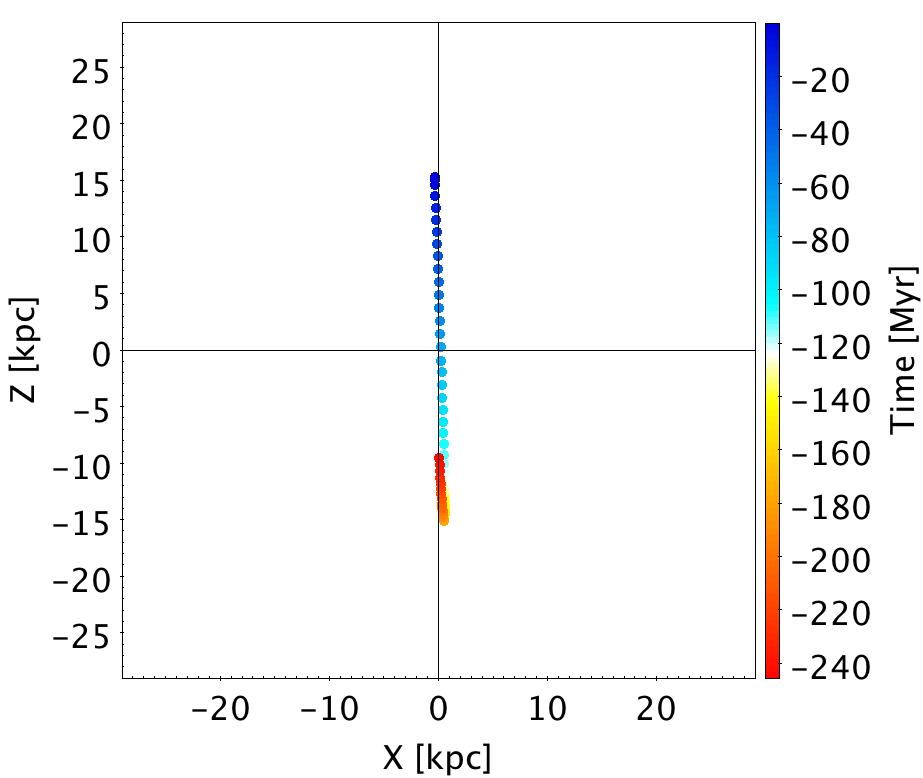}
	\includegraphics[width=\columnwidth]{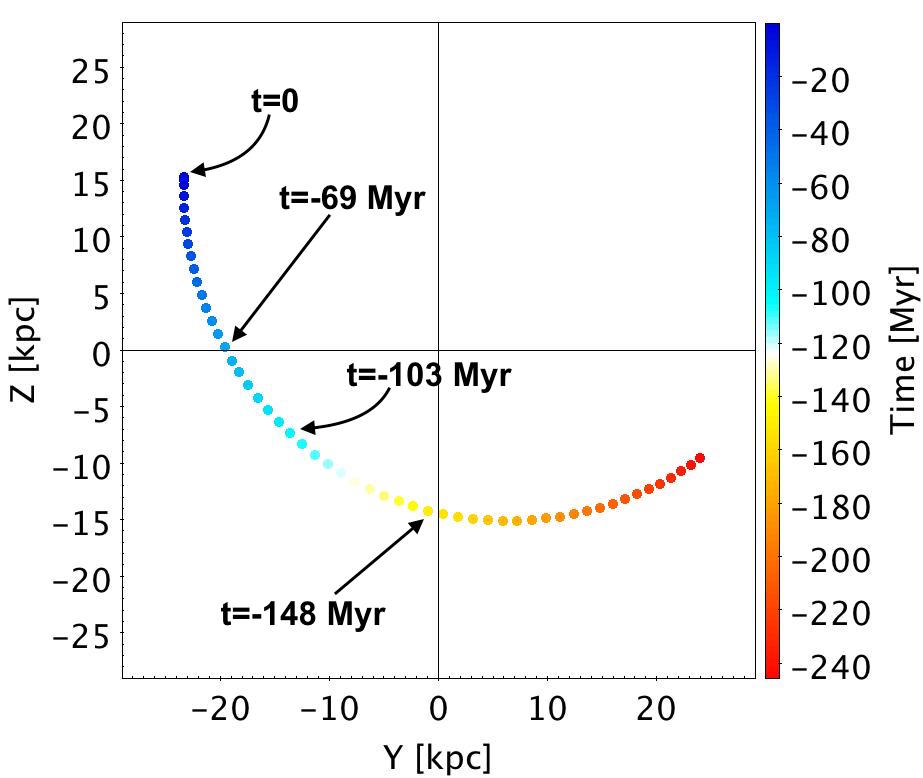}
    \caption{Orbit of PW~1 during the last 250~Myr projected into the X-Z (upper panel) and Y-Z (lower panel) planes of a right-handed Galactocentric Cartesian coordinate system. In this system the Sun is located at (X,Y,Z)=(-8.1,0,0). The epoch of each point in the orbit is color coded with the look-back time in Myr. In the lower panel we have labeled the points corresponding to a few remarkable epochs: the present day, corresponding to the current position of PW~1, the first computed position after the crossing of the Galactic Plane, 69~Myr ago, and the position of the system at the computed points nearest to 100 and 150 { Myr} ago.
    }
    \label{orbit}
\end{figure}

Adopting the position and mean proper motion from P19 and the radial velocity derived here, we integrated the orbit of PW~1 backward for 1.0~Gyr using GravPot16 \citep[][]{gravpot}\footnote{\tt https://fernandez-trincado.github.io/GravPot16/index.html}. This tool adopts the Galactic potential of the Besancon Galactic Model \citep{besa03}\footnote{\tt https://model.obs-besancon.fr}, including a rotating bar and an isothermal dark matter halo, and it has been recently used, within the Gaia Collaboration, to compute the orbits of many Galactic satellites \citep{gaia_hel}. We independently checked the results with other orbit integrators adopting different models of the Galactic potential, and we verified that the main results presented in this section are robust to these systematic uncertainties.

The path of the newly computed orbit during the last 250~Myr is displayed in Fig.~\ref{orbit}. The orbit is confined within $\la 1.5$~kpc of the Y-Z plane, with an inclination of $87.6\degr$ to the Galactic Plane. The perigalactic and apogalactic distances are $R_{peri}=13.8$~kpc and $R_{apo}=32.8$~kpc, eccentricity $e=0.41$, and period P=0.47~Gyr. The crossing of the Galactic disc occurs $t_{c}$=70~Myr ago. 
To explore the effect of the uncertainties of the initial conditions on the estimate of $t_{c}$ we repeated the integration over a coarse grid where, at each node, one of the relevant parameters 
($\langle \mu_{\alpha}\rangle$, $\langle \mu_{\delta}\rangle$, $\langle V_r\rangle$, and the distance)
was changed by $\pm 1\sigma$ and all the others were kept fixed at their best-estimated value. To be conservative, for the distance we adopted a $1\sigma$ uncertainty of 1.0~kpc, i.e. 10 times larger than that reported by PW1. Moreover, since the distance and the radial velocity are the parameters bearing the largest uncertainties and having the highest impact on $t_{c}$, we included in the grid also all the combinations implying simultaneous $1\sigma$ variations of both parameters. In total we explore twelve sets of initial conditions in addition to the best ones and we found that the impact on the predicted epoch of disc crossing 
is remarkably small, $68 \le t_{c}\le 74$~Myr. These results are summarised in Table~\ref{tab3}. 

\begin{table}
	\centering
	\caption{Times since the crossing of the Galactic Plane as a function of initial conditions for orbit integration}
	\label{tab3}
	\begin{tabular}{ccccc} 
		\hline
D   & $\langle \mu_{\alpha}\rangle$ & $\langle\mu_{\delta}\rangle$ & $\langle V_r\rangle$ & $t_c$\\
 ${\rm [kpc]}$ & [mas/yr] & [mas/yr] & [km/s] & [Myr]\\
		\hline
  28.9 & -0.56 &  0.47 &  275 &  70.1 \\
  27.9 & -0.56 &  0.47 &  265 &  69.6 \\
  27.9 & -0.56 &  0.47 &  275 &  68.0 \\
  27.9 & -0.56 &  0.47 &  285 &  68.6 \\
  28.9 & -0.52 &  0.47 &  275 &  69.8 \\
  28.9 & -0.56 &  0.47 &  265 &  71.8 \\
  28.9 & -0.56 &  0.47 &  285 &  68.6 \\
  28.9 & -0.56 &  0.49 &  275 &  69.5 \\
  28.9 & -0.56 &  0.45 &  275 &  70.8 \\
  28.9 & -0.60 &  0.47 &  275 &  70.5 \\
  29.9 & -0.56 &  0.47 &  265 &  73.9 \\
  29.9 & -0.56 &  0.47 &  275 &  72.2 \\
  29.9 & -0.56 &  0.47 &  285 &  70.7 \\
\hline
\multicolumn{5}{l}{The crossing time has been derived by linear unterpolation} \\
\multicolumn{5}{l}{between the two computed points of the orbit bracketing Z=0.0~kpc} \\
\multicolumn{5}{l}{The first row correspond to the best-estimate initial conditions.} \\
	\end{tabular}
\end{table}

Compared at face value with the age of the system derived above (100-150~Myr), $t_{c}\simeq 70_{-2}^{+4}$~Myr would imply that the star formation was ignited before the disc-crossing, when PW~1 was $\sim 7-14$~kpc below the plane. However, given the significant uncertainties that are still involved into this comparison, in our view, the similarity of the two timescales remains strongly suggestive of a connection between the crossing of the disc and the onset of star formation in PW~1. Presumably future data releases of Gaia and spectroscopic estimates of the metal content will significantly reduce the uncertainty in the orbit (distance and 3D motion) and in the age of the system, allowing us to establish if these key events have been simultaneous or not. 

\subsection{The parent gas cloud of PW~1}

PW~1 is a very special case among the stellar systems orbiting the Milky Way. In particular it is the only known object lying so far from the Galactic Disc (Z$\simeq 15$~kpc) that is made only of young stars (age$\la 150$~Myr), calling for recent association with some \ion{H}{i} structure/cloud. Neutral hydrogen is also relatively rare in these remote regions of the Galactic Halo, hence the proximity of PW~1 to { an arm} of the main \ion{H}{i} structure within the Halo, the Magellanic Stream, is strongly suggestive of a physical association. 
Small-scale episodes of recent star formation likely associated to the interactions between the two Magellanic Clouds and the Milky Way are known to occur, e.g. in the Magellanic Bridge, in the outskirts of the Small Magellanic Cloud \citep{bridge,david}, and possibly also in the LA \citep{dana}\footnote{ However, \citet{zhang} showed that the orbital parameters of most of the Casetti-Dinescu et al. stars are not compatible with an association with the LA. Moreover, we verified that none of the stars followed-up by \citet{zhang} is compatible with the orbit of PW~1. In particular, all of them have $Z_{max}< 10.0$~kpc, to be compared to $Z_{max}= 32.6$~kpc of our fiducial orbit of PW~1 and to its current height above the Galactic plane, $Z=15.2$~kpc.}

P19 suggested that the crossing of the Galactic Disc by the head of the MS { Leading Arm} triggered the star formation episode that produced PW~1. Then, the newly formed stars decoupled from the parent \ion{H}{i} cloud, as they are free from the drag exerted on the moving cloud by the Galactic gas in the disc and in the corona.
If this were to be the case, a velocity and position lag between PW~1 and the parent gas structure may be expected. 
However since both the star formation event and the crossing of the disc are recent (70-150~Myr), it may be expected that the lag should be small, and the nearest \ion{H}{i} in projection should also display the smallest velocity lag.

\begin{figure*}
	\includegraphics[width=\textwidth]{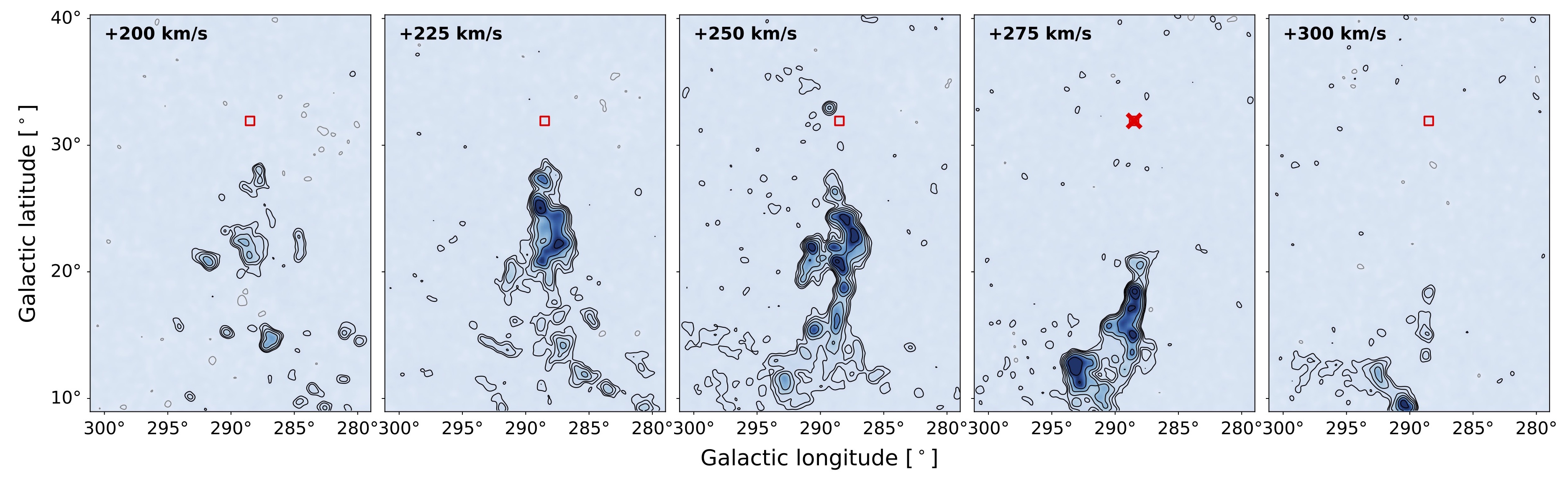}
    \caption{\ion{H}{i} density fields from HI4PI mapping the region of the edge of the MS Leading Arm near PW~1, in five radial velocity channels bracketing the velocity of the stellar system. The adopted spatial resolution is $0.54\degr$ FWHM. The contours correspond to brightness temperatures ranging from $T_B=$~0.035~K (corresponding to $3\times$ the rms noise), to 2.24~K, increasing in steps of factor of 2.
    An empty red box marks the position of PW~1 in all the maps, an additional red cross highlights the map of the Vr=+275~km/s channel, corresponding to the velocity of PW~1.
    }
    \label{channels}
\end{figure*}

In Fig.~\ref{channels} we show a series \ion{H}{i} ~channel maps in the surroundings of PW~1, and focused on the MS { Leading Arm}, from the all-sky HI4PI survey \citep{HI4PI}.
This figure shows that (a) indeed, there is no \ion{H}{i} with $T_B\ge 0.035$~K that coincides in position with PW~1, and (b) the \ion{H}{i} velocity trend is in the opposite sense with respect to that envisaged above. 
As the radial velocity of the gas increases from +200~km/s to reach the velocity of the stars (+275~km/s) and beyond, the mean latitude of the gas distribution decreases, moving away from the position of PW~1 toward the Galactic Disc.
While the Northern edge of the gas distribution at $V_r=200-225$~km/s is about $\sim 4\degr$ apart from the stellar system, corresponding to $\sim 2$~kpc at D=28.9~kpc, at $V_r=275$~km/s the gas is $\sim 10\degr$ apart, corresponding to $\sim 5$~kpc\footnote{Note that even if the radial velocity is the same, the 10 degrees distance in the sky implies also a difference of $\sim 20$~km/s in $V_{GSR}$.}. The described trend is in good agreement with the overall gradient observed along the LA by \citet{venz}, with the gas decelerating from $V_{GSR}=84.2$~km/s, at the position of the Large Magellanic Cloud, to about $V_{GSR}=6$~km/s in the proximity of PW~1, that, instead, has $V_{GSR}=84\pm 10$~km/s. 

{ Realistic hydrodynamical modelling of the complex interaction between the LA and the Galactic gaseous disc and corona are required to explore in detail the evolutionary paths that may have led to the observed configuration. Below we briefly discuss two possible scenarios that are broadly compatible with the available data.}

\subsubsection{Dissolution of the parent cloud}

The lack of a good match with known structures 
may suggest that the parent cloud of PW~1 is not detectable { in \ion{H}{i}} anymore, having already disappeared.
Indeed, fragmentation is observed to occur in the MS \citep[][and references therein]{nide08,nide10,for}, especially at the edge of the Leading Arm, near PW~1 \citep{for,venz,dong}, and low mass clouds of cold gas { ($M\le 10^5~M_{\sun}$)} may be dissolved on short time-scales ($\la 100-200$~Myr, depending on the initial mass) by the interaction with the  circum-galactic medium \citep[see, e.g.][and references therein]{hp,armi}. { In particular, \citet{tepperSC}, based on the results of a set of hydrodynamical simulations of the evolution of the Smith cloud \citep{smith}, concluded that a massive ($M\ga 10^8~M_{\sun}$), dark-matter free High Velocity Cloud \citep[HVC,][]{WW} crossing the Galactic disc with Galactocentric distance and velocity similar to PW~1, cannot survive the transit.}

The signatures of on-going ram pressure stripping on LA structures have been noted and discussed by \citet{venz}. 
In the present case, the feedback from star formation (supernovae) could have contributed to blow the residual gas away from the stellar system. In this context it is { interesting} to note the compact cloudlet at (l,b)$\simeq(289.4\degr,32.9\degr)$ in the $V_r=250$~km/s snapshot of Fig.~\ref{channels}. This is the known \ion{H}{i} structure nearest to PW~1, in projection (just $\simeq 1.3\degr$, corresponding to $\simeq 0.7$~kpc), with a velocity difference that is relatively modest (21.5~km/s, in the Galactic standard of rest). At the distance of PW~1 its integrated flux corresponds to an \ion{H}{i} ~mass $M_{HI}\simeq 3000\pm 300$~M$_{\sun}$, about twice the total stellar mass of PW~1, hence it can be a plausible candidate for the gaseous residual of the formation of PW~1.

\subsubsection{Flying away from HVC~287.5+22.5+240}

{ 
On the other hand, Fig.~\ref{channels} may be seen, from right to left, as depicting the progressive deceleration of the gas cloud that gave origin to PW~1 as it penetrates in the northern galactic hemisphere after having crossed the disc, while PW~1 is flying away, free of any drag from the Galactic gas. Note that this may happen independently of the actual origin of the LA \citep[gas tidally + ram-pressure stripped from the SMC or trailing gas from a LMC/SMC satellite running ahead of the MCs, as suggested, e.g., by][]{tepper}. In this scenario the clouds nearest to PW~1 are those that have suffered the highest degree of deceleration, thus showing a larger velocity lag with respect to the stellar system. 

In this context, it is interesting to note that the gas structure that dominates the map in the central panel of Fig.~\ref{channels}, at $V_r=+250$~km/s (corresponding to $V_{LSR}\simeq 245$~km/s), is a well known and thoroughly studied HVC: HVC~287.5+22.5+240 \citep[also known as WW~187, see][and references therein]{wakker}. In particular, detailed chemical abundance analyses have been performed by various authors, from UV spectra of the background Seyfert galaxy NGC~3783 \citep{lu,wakker,fox,richter}. The abundance pattern is found to match very closely that of the SMC, thus pointing to an origin from an episode of gas stripping from this galaxy \citep[][their Fig. 6, in particular]{richter}.

It is very interesting to note that the abundance of sulphur in HVC~287.5+22.5+240, an element that is virtually immune from dust depletion, is $[S/H]\simeq -0.6$ \citep{lu,wakker,richter}. Adopting $[S/Fe]\simeq +0.6$ for SMC stars, according to \citet{RD92} and following \citet{lu}, and assuming a common original composition for the HVC and the SMC, $[Fe/H]\simeq -1.2$ is obtained for HVC~287.5+22.5+240, remarkably similar to the available estimate for PW~1 ($[Fe/H]=-1.1$).

Moreover, \citet{mcg} found that HVC~287.5+22.5+240 is the only case, among the 27 HVCs/gas complexes analysed by these authors, displaying a strong and coherent magnetic field. According to \citet{mcg}, the observed field is consistent to that required to dynamically stabilise the cloud against ram pressure. This factor may have played
a major role to let HVC~287.5+22.5+240 survive the crossing of the Galactic disc, providing a natural way out from the conclusions by \citet{tepperSC}.
Finally, molecular hydrogen, typically conducive to star formation, has been detected in this HVC \citep{sem,richter}.

These evidences provide significant support to the hypothesis that HVC~287.5+22.5+240 is the parent cloud of PW~1, within the interpretative scheme originally advanced by P19 for the origin of PW~1.

}

\section{Summary and conclusions}
\label{conclu}

We have acquired { EFOSC2@NTT} medium-resolution spectra of five members of the recently discovered young stellar system Price-Whelan~1, lying at D$\simeq 29$~kpc from us, at an height of $\simeq 15$~kpc from the Galactic plane.
These spectra allowed us to obtain the first { measurement} of its systemic radial velocity $V_r=275 \pm 10$~km/s. Having at disposal all the three components of the spatial velocity of PW~1 we computed its orbit within a realistic Galactic potential. According to the newly derived orbit the system crossed the Galactic plane 70~Myr ago, a timescale comparable with its estimated age (100-150~Myr).

The proximity with the Leading Arm of the Magellanic Stream lead to the hypothesis that PW~1 formed at the edge of this huge \ion{H}{i} structure (P19). However its radial velocity is significantly different from the gas in the nearest edge of the LA. We briefly discuss the possibility that the parent cloud of PW~1 has already dissolved into the Galactic corona { and also the possible identification of the nearby high velocity cloud HVC~287.5+22.5+240 as the birthplace of PW~1}.

We show that PW~1 is made of three main pieces (plus additional likely members dispersed in the surroundings), and we provide estimates of the  stellar mass, absolute integrated V magnitude, and a proxy of the half light radius for each of them. We provide also evidence that the stellar content of the three pieces is not exactly the same, the brightest / most massive stars residing almost exclusively in the aE subsystem. 

A direct spectroscopic metallicity estimate is probably the most relevant piece of information that is currently missing to constrain the nature and the origin of this very unusual stellar system, as it would strongly mitigate the effects of the age/distance/metallicity degeneracy that now is leaving room to significant systematic uncertainties in these key parameters. Resolving the internal kinematics would also be very useful to finally establish if PW~1 is indeed dissolving, also constraining the relevant timescales for this process.

\section*{Acknowledgements}

We are grateful to an anonymous referee for providing very useful suggestions that significantly improved this paper.
M.B. is grateful to A. Sollima, F. Fraternali, A. Bragaglia, D. Romano and A. Mucciarelli for useful suggestions and discussion, and to J.G. Fern\'andez-Trincado for his precious help with GalPot16.

Based on observations collected at the European Southern Observatory under ESO programme 103.B-0568(A).

This project has received funding from the European Research Council (ERC) under the European Union's Horizon 2020 research and innovation programme (grant agreement No. 834148)

This work has made use of data from the European Space Agency (ESA) mission Gaia
(http://www.cosmos.esa.int/gaia), processed by the Gaia Data Processing and
Analysis Consortium (DPAC, http://www.cosmos.esa.int/web/gaia/dpac/consortium).
Funding for the DPAC has been provided by national institutions, in particular
the institutions participating in the Gaia Multilateral Agreement.

This work has made use of data from the Pan-STARRS1 Surveys.
The Pan-STARRS1 Surveys (PS1) and the PS1 public science archive have been made possible through contributions by the Institute for Astronomy, the University of Hawaii, the Pan-STARRS Project Office, the Max-Planck Society and its participating institutes, the Max Planck Institute for Astronomy, Heidelberg and the Max Planck Institute for Extraterrestrial Physics, Garching, The Johns Hopkins University, Durham University, the University of Edinburgh, the Queen's University Belfast, the Harvard-Smithsonian Center for Astrophysics, the Las Cumbres Observatory Global Telescope Network Incorporated, the National Central University of Taiwan, the Space Telescope Science Institute, the National Aeronautics and Space Administration under Grant No. NNX08AR22G issued through the Planetary Science Division of the NASA Science Mission Directorate, the National Science Foundation Grant No. AST-1238877, the University of Maryland, Eotvos Lorand University (ELTE), the Los Alamos National Laboratory, and the Gordon and Betty Moore Foundation.

Most of the analysis presented in this paper has been performed with TOPCAT \citep{topcat}.
This research made use of the GravPot16 software, a community-developed core under the git version-control system on GitHub.
This research has made use of the SIMBAD database, operated at CDS, Strasbourg, France.
This research has made use of the NASA/IPAC Extragalactic Database (NED) which is operated by the Jet Propulsion Laboratory, California Institute of Technology, under contract with the National Aeronautics and Space Administration. 
This research has made use of NASA's Astrophysics Data System.





\begin{thebibliography}{999}
\bibitem[Armillotta et al.(2017)]{armi}
         Armillotta, L., Fraternali, F., Werk, J.K., Prochaska, J.X., Marinacci, F.,
         2017, \mnras, 470, 114 
\bibitem[Babusiaux et al.(2018)]{gaia_babu}
         Gaia Collaboration, Babusiaux, C., et al., 2018, \aap, 616, A10 
\bibitem[Bellazzini et al.(2006)]{bellaz06}
         Bellazzini, M., Ibata, R., Martin, N., Lewis, G. F., Conn, B., Irwin, M. J., 2006, \mnras, 366, 865
\bibitem[Bellazzini et al.(2018)]{alone}
         Bellazzini, M., Armillotta, L., Perina, S., et al., 2018, \mnras, 476, 4565
\bibitem[Ben Bekhti et al.(2016)]{HI4PI}
         Ben Bekhti, N., et al. 2016, \aap, 594, A116 
\bibitem[Bressan et al.(2012)]{bressan}
         Bressan, A., Marigo, P., Girardi, L., Salasnich, B., Dal Cero, C., 
         Rubele, S., Nanni, A., 2012, \mnras, 427, 127 
\bibitem[Butler et al.(2007)]{butler}
         Butler, D.J., Mart\'inez-Delgado, D., Rix, H.-W., Pe\~narrubia, J., 
         de Jong, J.T.A., 2007, \aj, 133, 2274
\bibitem[Casetti-Dinescu et al.(2014)]{dana}
         Casetti-Dinescu, D.I., Moni Bidin, C., Girardi, R.M., et al., 2014, \apj, 784, L37
\bibitem[Chambers et al.(2016)]{pan}
         Chambers, K.C., et al., 2016, arXiv:1612.05560
\bibitem[D'Onghia \& Fox(2016)]{dong}
         D'Onghia, E., Fox, A.J., 2016, \araa, 54, 363
\bibitem[Evans et al.(2018)]{gaia_eva}
         Evans, D.W., et al., 2018, \aap, 616, A4
\bibitem[Brown et al.(2018)]{gaia_bro}
         Gaia Collaboration, Brown, A., et al., 2018, \aap, 616, A1
\bibitem[Cantat-Gaudin et al.(2018)]{CG18}
         Cantat-Gaudin, T., Jordi, C., Vallenari, A., et al., 2018, \aap, 618, 93
\bibitem[Cantat-Gaudin et al.(2019)]{CG19}
         Cantat-Gaudin, T., Krone-Martins, A., Sedaghat, N., et al., 2019, \aap, 624, A12
\bibitem[Fern\'andez-Trincado et al.(2017)]{gravpot}
         Fern\'andez-Trincado, J.G., Robin, A.C., Moreno, E., P\'erez-Villegas, A.,
         Pichardo, B., in SF2A-2017: Proceedings of the Annual meeting of the French 
         Society of Astronomy and Astrophysics, C. Reyl\'e et al. Eds., p. 193
\bibitem[For et al.(2014)]{for}
         For, B.Q., Staveley-Smith, L., Matthews, D., McClure-Griffiths, N.M., 
         2014, \apj, 792, 43
\bibitem[Fox et al.(2018)]{fox}
         Fox, A.J., Barger, K.A., Wakker, B.P., et al., 2018, \apj, 854, 142 
\bibitem[Hammer et al.(2015)]{hammer}
         Hammer, F., Yang, Y.B., Flore, H., Puech, M., Fouquet, S., 2015, \apj, 813, 110
\bibitem[Harris (1996)]{h96}         
         Harris, W.E. 1996, \aj, 112, 1487
\bibitem[Heitsch \& Putman(2009)]{hp}
         Heitsch, F., \& Putman, M.E., 2009, \apj, 698, 1485 
\bibitem[Helmi et al.(2018)]{gaia_hel}
         Gaia Collaboration, Helmi, A., et al., 2018, \aap, 616, A12
\bibitem[Helmi et al.(2018)]{gaia-ence}
         Helmi, A., Babusiaux, C., Koppelman, H.H., Veljanoski, J., Brown, A.G.A., 2018, Nature, 563, 85
\bibitem[Ibata et al.(1994)]{sgr0}
         Ibata, R. A., Gilmore, G.,\& Irwin, M. J. 1994, Nature, 370, 194
\bibitem[Ibata et al.(2019)]{gap_abyss}
         Ibata, R.A., Malhan, K., Martin, N.F., 2019, \apj, 872, 152
\bibitem[Lindgren et al.(2018)]{gaia_lin}
         Lindgren, L., et al., 2018, \aap, 616, A2
\bibitem[Lu et al.(1998)]{lu}
         Lu, L., Savage, B.D., Sembach, K.R., Wakker, B.P., Sargent, W.L.W.,
         Oosterloo, T.A., 1998, \aj, 115, 162
\bibitem[Malhan et al.(2018)]{ghostly}
         Malhan, K., Ibata, R.A., Martin, N.F., 2018, \mnras, 481, 3442
\bibitem[Majewski et al.(2003)]{m03}         
         Majewski, S. R., Skrutskie, M. F., Weinberg, M. D., \& Ostheimer, J. C. 2003, \apj, 599, 1082
\bibitem[Marigo et al.(2017)]{marigo}
         Marigo, P., Girardi, L.Bressan, A., et al., 2017, \apj, 835, 77
\bibitem[McMillan(2017)]{McM}
         McMillan, P.J., 2017, \mnras, 465, 76
\bibitem[Martin et al.(2004)]{mart04}
         Martin, N.F., Ibata, R.A., Bellazzini, M., Irwin, M. J., Lewis, G. F., Dehnen, W., 2004, 348, 12
\bibitem[Martin et al.(2018)]{mart18}
         Martin, N.F., Collins, M.L.M., Longeard, N., Tollerud, E., 2018, \apj, 859, L5
\bibitem[Martinez-Delgado et al.(2019)]{david}
         Martinez-Delgado, D., et al., \aap, submitted (arXiv:1907.02264)
\bibitem[Massey et al.(2006)]{m33}
         Massey, P., Olsen, K.A.G., Hodge, P.W., Strong, S.B., Jacoby, G.H., Schlingman, W., Smith, R.C., 2006, \aj, 131, 2478
\bibitem[McClure-Griffiths et al.(2010)]{mcg}
         McClure-Griffiths, N.M., Madsen, G.J., Gaensler, B.M, McConnel, D., 
         Schnitzler, D.H.F.M., 2010, \apj, 725, 275
\bibitem[McConnachie(2012)]{mc12}
         McConnachie, A.W., 2012, \aj, 144, 4
\bibitem[Momany et al.(2006)]{moma}
         Momany, Y., et al., 2006, \aap, 451, 515
\bibitem[Monaco et al.(2003)]{sgrbhb}
         Monaco, L., Bellazzini, M., Ferraro, F.R., Pancino, E., 2002, \apj, 597, L25
\bibitem[Montenegro et al.(2019)]{bulgehb}
         Montenegro, K., Minniti, D., Alonso-Garc\'ia, J., Hempel, M., Saito, R.K., Beers, T.C., Brown, D., 
         2019, \apj, 872, 206
\bibitem[Nidever et al.(2008)]{nide08}
         Nidever, D. L., Majewski, S. R., Butler Burton, W., 2008, ApJ, 679, 432
\bibitem[Nidever et al.(2010)]{nide10}
         Nidever, D. L., Majewski, S. R., Butler Burton, W., Nigra, L. 2010, \apj, 723, 1618,
\bibitem[Price-Whelan et al.(2019)]{pw1}
         Price-Whelan, A.M., Nidever, D.L., Choi, Y., Schlafly, E.F., Morton, S.E., Belokurov, V., 2019, \apj, 
         in  press (arXiv:1811.05991) 
\bibitem[Putman et al.(1998)]{mary}
         Putman, M.E,, Gibson, B.K., Staveley-Smith, L, et al. 1998, Nature, 394, 752
\bibitem[Richter et al.(2018)]{richter}
         Richter, P., Fox, A.J., Wakker, B.P., et al., 2018, \apj, 865, 145
\bibitem[Robin et al.(2003)]{besa03}
         Robin, A.C., Reyl\`e, C., Derri\`ere, S., Picaud, S., 2003, \aap, 409, 523 
\bibitem[Russel \& Dopita(1992)]{RD92}
         Russell, S. C., \& Dopita, M. A. 1992, \apj, 384, 508
\bibitem[Sand(2017)]{sand17}
         Sand D. J. et al., 2017, \apj, 843, 134
\bibitem[Sembach et al.(2001)]{sem}
         Sembach, K.R., Howk, J.C., Savage, B.D., Schull, J.M., 2001, \aj, 121, 992
\bibitem[Schlafly \& Finkbeiner (2011)]{sf11}	     
         Schlafly, E.F. \& Finkbeiner, D.P., 2011, \apj, 737, 103 
\bibitem[Schlegel et al. (1998)]{sfd}
         Schlegel, D. J., Finkbeiner, D. P., \& Davis, M. 1998, \apj, 500, 525
\bibitem[Sch\"onrich, Binney \& Dehnen(2010)]{schon}
         Sch\"onrich, R., Binney, J., Dehnen, W., 2010, \mnras, 2010, 403, 1829
\bibitem[Skowron et al.(2014)]{bridge}
         Skowron, D.M., et al., 2014, \apj, 795. 108 
\bibitem[Smith(1963)]{smith}
         Smith G. P., 1963, Bull. Astron. Inst. Neth., 17, 203
\bibitem[Taylor(2005)]{topcat}
         Taylor, M.B., 2005, in Astronomical Data Analysis Software and Systems XIV, 
         ASP Conference Series, Vol. 347,  P. Shopbell, M. Britton, and R. Ebert Eds. 
         San Francisco: Astronomical Society of the Pacific,  p.29 
\bibitem[Tepper-Garc\'ia \& Bland-Hawthorn(2018)]{tepperSC}
         Tepper-Garc\'ia, T., Bland-Hawthorn, 2018, \mnras, 473, 5514
\bibitem[Tepper-Garc\'ia et al.(2019)]{tepper}
         Tepper-Garc\'ia, T., Bland-Hawthorn, J., Pawlowski, M.S., Fritz, T.K., 
         2019, \mnras, 488, 918
\bibitem[Torrealba et al.(2019)]{antlia2}
         Torrealba, G., Belokurov, V., Koposov, S.E., et al., 2019, \mnras, in press (arXiv:1811.04082)
\bibitem[Venzmer, Kerp \& Kalberla(2012)]{venz}
         Venzmer, M.S., Kerp, J., Kalberla, P.M.V., 2012, \aap, 547, A12
\bibitem[Wakker \& van Woerden(1997)]{WW}
         Wakker, B.P., van Woerden, H., 1997, \araa, 35, 217
\bibitem[Wakker et al.(2002)]{wakker}
         Wakker, B.P., Oosterloo, T.A., Putman, M.E., 2002, \aj, 123, 1953
\bibitem[Wang et al.(2019)]{wang}
         Wang, J., Hammer, F., Yang, Y., Ripepi, V., Cioni, M.-R., 
         Puech, M., Flores, H., 2019, \mnras, 486, 5907
\bibitem[Zhang et al.(2019)]{zhang}
         Zhang, L., Casetti-Dinescu, D.I., Moni Bidin, C., et al., 2019, \apj, 871, 99
\end{thebibliography}








\bsp	
\label{lastpage}
\end{document}